\documentstyle[12pt,epsf]{article}

\newcommand{\eq}{\begin{equation}}
\newcommand{\en}{\end{equation}}

\newcommand{\eqa}{\begin{eqnarray}}
\newcommand{\ena}{\end{eqnarray}}
\newcommand{\eqan}{\begin{eqnarray*}}
\newcommand{\enan}{\end{eqnarray*}}

\newcommand{\lbl}{\label}

\newcommand{\sect}[1]{\setcounter{equation}{0}\section{#1}}

\newcommand{\AP}[1]{Ann. Phys.\ {\bf #1}\ }

\newcommand{\NP}[1]{Nucl. Phys.\ {\bf #1}\ }

\def\sqr#1#2{{\vcenter{\hrule height.#2pt
     \hbox{\vrule width.#2pt height#1pt \kern#1pt
        \vrule width.#2pt}
     \hrule height.#2pt}}}

\def\thinspace{\kern .16667em}

\def\Dir{\nabla\kern-2ex\Big{/}}

\def\Dsl{\partial\kern-1.5ex\Big{/}}
\def\slp{{p\kern-1ex {/}}}
\def\slq{{q\kern-1ex {/}}}
\def\slk{{k\kern-1.2ex {/}}}
\def\sle{{\epsilon \kern-1ex {/}}}
\def\slb{{b \kern-1ex {/}}}

\def\sq2{\sqrt{2}}

\def\reali{{\hbox{\s@ l\kern-.5ex R}}}
\def\naturali{{\hbox{\s@ l\kern-.5ex N}}}
\def\interi{{\mathchoice
 {\hbox{Z\kern-1.5mm Z}}
 {\hbox{Z\kern-1.5mm Z}}
 {\hbox{{Z\kern-1.2mm Z}}}
 {\hbox{{Z\kern-1.2mm Z}}}  }}

\def\unity{{\hbox{\s@ 1\kern-.8mm l}}}
\def\uno{{\hbox{ 1\kern-.8mm l}}}
\def\part{\partial}

\def\aa{\alpha}

\def\dd{\delta}

\def\gg{\gamma}
\def\GG{\Gamma}

\begin{document}
\begin{titlepage}
\begin{flushright}
Bologna preprint\\
DFUB-95-18\\
September 1995\\
\end{flushright}
\vspace*{0.5cm}
\begin{center}
\begin{Large}
{\bf GAUGE INVARIANCE ON BOUND STATE ENERGY LEVELS
\footnote{Talk given at the NATO-ASI, {\it Electron Theory 
and Quantum Electrodynamics}, Edirne 8-16 September 1994}
\\}
\end{Large}
\vspace*{1.5cm}
         {{\large Antonio Vairo}\footnote
{E-mail: vairo@bo.infn.it}}
         \\[.3cm]
          I.N.F.N. - {\it Sez. di Bologna and Dip. di Fisica,
         \\[.1cm]
          Universit\`a di Bologna, Via Irnerio 46, I-40126 Bologna,
          Italy}\\
\end{center}
\vspace*{0.7cm}
\begin{abstract}
{
In this paper the problem of the gauge in a bound state calculation 
is discussed. In particular, in order to verify the gauge invariance 
in the energy levels expansion, some set of gauge invariant contributions 
are given.
}
\end{abstract}
\vfill
\end{titlepage}

\setcounter{footnote}{0}
\def\ut{{\tilde u}}
\def\zt{{\tilde z}}
\def\dz{{\sqrt{2}z}}

\def\uij{U_{ij}}
\def\ucij{U^\dagger_{ij}}
\def\uji{U_{ji}}
\def\ucji{U^\dagger_{ji}}
\def\dag{\dagger}
\def\zpm{z_{\pm}}
\def\zp{z_+}
\def\zm{z_-}
\def\ddt{{\dd T}}
\def\mucr{\mu_{cr}}

\newcommand{\mat}[4]{\left( 
                     \begin{array}{cc}
                     {#1} & {#2} \\
                     {#3} & {#4} 
                     \end{array}
                     \right)
                    }

\newcommand{\ft}[3]{ {d^{#1}{#2}\over (2\pi)^{#1}} ~ e^{i {#2}\cdot{#3}} }
\newcommand{\ftt}[2]{{d^{#1}{#2}\over (2\pi)^{#1}}}

\def\rhm{\rho_-}
\def\rhp{\rho_+}
\def\sgm{\sigma_-}
\def\sgp{\sigma_+}

\def\rd{\sqrt{2}}
\def\usrd{{1\over\sqrt{2}}}
\def\dxy{\delta^2(x-y)}
\def\dij{\delta^{ij}}
\def\dsi{\partial_{x^-}}
\def\dta{\partial_{x^+}}

\newcommand\modu[1]{|{#1}|}
\newcommand\fai[4]{{#1}^{{#2}}_{{#3}} ({#4}) }

\def\ggv{{\vec \gg}}
\def\pv{{\vec p}}
\def\qv{{\vec q}}
\def\kv{{\vec k}}

\sect{Introduction}

In this paper I review some basics concerning the gauge invariance 
on bound state ~\cite{ffh}.

In the first section some simple examples about gauge invariance 
on mass-shell are given. In the second section there are some 
general considerations about gauge invariance in a off shell problem
and up to order $\aa^4$ the calculation of some contributions
to the energy levels of positronium in Feynman gauge. 
The full cancellation of the spurious $\aa^3 \log \aa$, $\aa^3$
terms which arise typically in this gauge, is performed.
The calculations are done in the Barbieri-Remiddi bound 
state formalism ~\cite{bar}, ~\cite{rem}. $K_c$, $\psi_c$ are 
the Barbieri-Remiddi zeroth-order kernel and the corresponding 
wave function.

\sect{Gauge invariance on mass-shell.}

It is generally easy to verify the gauge invariance in a scattering process.
The incoming and outcoming particles are on mass-shell and the related 
wave functions are not $\alpha$-depending. This involves that each Feynman 
graph contributes to the scattering amplitude only at the order 
in the fine structure constant determined by his number of vertices. 
In the following I will verify the gauge invariance at the leading 
order in $\alpha$ in two very simply cases: the Compton and the 
$e^+ ~ e^-$ scattering.

At the order $\alpha$  only the two graphs of Fig.~\ref{pscompton}
contribute to the Compton scattering amplitude.
\begin{figure}[htb]
\vskip 0.8truecm
\makebox[2.2truecm]{\phantom b}
\epsfxsize=9truecm
\epsffile{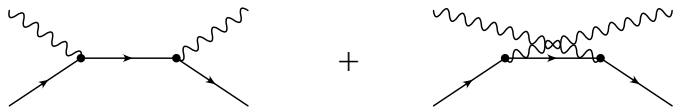}
\vskip 0.3truecm
\caption{{{\small  Compton scattering at the tree level.}}}
\lbl{pscompton}
\vskip 0.8truecm
\end{figure}
\\ The contribution to the scattering amplitude coming from the graphs of 
Fig.~\ref{pscompton} is:
\eqa
A_{Com}(p_1,k_1;p_2,k_2) &=& 
- (2\pi)^4 \delta^{(4)}(p_1+k_1 - p_2 - k_2)~\aa
\lbl{com}\\
&\cdot& \Big(
\bar u (p_2) \sle (k_2) S_F(p_1+k_1) \sle (k_1) u(p_1) 
\nonumber\\
&+& 
~\bar u (p_2) \sle (k_1) S_F(p_1-k_2) \sle (k_2) u(p_1) \Big ) ~,
\nonumber
\ena 
where $k_1$, $p_1$ and $k_2$, $p_2$ are the incoming and outcoming momenta 
(on mass shell: $p_j^2 = m^2$ for the electron momenta and 
$k_j^2 = 0$ for the photon momenta), $\epsilon^\mu$ is the photon 
polarization, $S_F$ the fermion free propagator:
\eq
S_F(p) = {i \over \slp -m +i\epsilon} ~,
\lbl{sf}
\en
and $u$, $\bar u$, $v$, $\bar v$ are the Dirac spinors ($(\slp-m)u(p) = 0$, 
$(\slp+m)v(p) = 0$).
\\ The gauge variation (e.g. on the second external photon) of $A_{Com}$ i.e.
$A_{Com}^{Gauge\#1} - A_{Com}^{Gauge\#2}$ is:
\eqa
\dd A_{Com}(p_1,k_1;p_2,k_2) 
&=& - (2\pi)^4 \delta^{(4)}(p_1+k_1 - p_2 - k_2)~\aa
\lbl{com2}\\
&\cdot&
\bar u (p_2) \Big \{ \slk_2 S_F(p_2+k_2) \sle (k_1) 
\nonumber\\
&+& \sle (k_1) S_F(p_1-k_2) \slk_2 \Big\} u(p_1) 
\nonumber\\
&=&-(2\pi)^4 \delta^{(4)}(p_1+k_1 - p_2 - k_2)~\aa
\nonumber\\
&\cdot&
\bar u (p_2)
\Big \{ \left ( i - (\slp_2 - m) S_F(p_2+k_2) \right ) \sle (k_1) 
\nonumber\\
&+& \sle (k_1) \left ( - i - S_F(p_1-k_2) (\slp_1 - m) \right ) \Big\} u(p_1) 
\nonumber\\
&=&-(2\pi)^4 \delta^{(4)}(p_1+k_1 - p_2 - k_2) ~\aa
\nonumber\\
&\cdot&
\bar u (p_2) \Big \{ i \sle(k_1) - i \sle(k_1) \Big \} u(p_1) ~ = ~0 ~,
\nonumber
\ena 
which verifies the gauge invariance.

The second example is the $e^+ ~ e^-$ scattering amplitude. At the order 
$\alpha$ contributions to the scattering amplitude arise only from the 
graphs of Fig.~\ref{psepem}.
\begin{figure}[htb]
\vskip 0.8truecm
\makebox[2.2truecm]{\phantom b}
\epsfxsize=9truecm
\epsffile{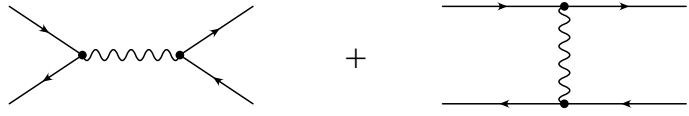}
\vskip 0.3truecm
\caption{{{\small  $e^+ ~ e^-$ scattering at the tree level.}}}
\lbl{psepem}
\vskip 0.8truecm
\end{figure}
The photon propagator $D_{\mu\nu}$ is gauge dependent; in Feynman and 
Coulomb gauge one can write it as:
\eqa
D^{Fey}_{\mu\nu}(p)  &=& -i{g_{\mu\nu} \over p^2+i\epsilon} ~;
\lbl{DFey}\\
D^{Cou}_{\mu\nu}(p)  &=& -i{1 \over p^2+i\epsilon} 
\left\{g_{\mu\nu} - {\eta\cdot p (p_\mu\eta_\nu+p_\nu\eta_\mu) - p_\mu p_\nu 
\over \pv^{~2}} \right\}
\nonumber\\
&~& \eta^\mu = (1,\vec 0~) ~.
\lbl{DCou}
\ena
The difference between the propagators in the above gauges can be expressed as:
\eqa
D^{Cou}_{\mu\nu}(p) - D^{Fey}_{\mu\nu}(p) &=&
b_\mu(p)p_\nu+b_\nu(p)p_\mu ~,
\nonumber\\
b_\mu(p) &\equiv& {i\over 2 (p^2+i\epsilon)} {1\over \vec p ^{~2}}
(-p_\mu+2\dd_{0\mu} p_0 ) ~.
\lbl{dD}
\ena
Not only the sum of the graphs in Fig.~\ref{psepem} is gauge invariant but 
each graph also. To verify the gauge invariance of the annihilation graph one
needs only equation (\ref{dD}) and the Dirac equation for the spinors
$u$ and $v$:
\eqa
\dd A_{ann}(p_1,k_1;p_2,k_2) &=& 
- (2\pi)^4 \delta^{(4)}(p_1+k_1 - p_2 - k_2) ~\aa
\lbl{epem}\\
&\cdot&
\bar v (k_1) \gamma^\mu u(p_1) \Big\{D^{Cou}_{\mu\nu}(p_1+k_1)
\nonumber\\
&-& 
D^{Fey}_{\mu\nu}(p_1+k_1) \Big\} 
\bar u (p_2) \gamma^\nu v(k_2) 
\nonumber\\
&=& 
- (2\pi)^4 \delta^{(4)}(p_1+k_1 - p_2 - k_2)~\aa
\nonumber\\
&\cdot& 
\bar v (k_1) \gamma^\mu u(p_1) \Big\{
b_\mu(p_1+k_1)~(p_{1\nu}+k_{1\nu}) 
\nonumber\\
&+& 
b_\nu(p_1+k_1)~(p_{1\mu}+k_{1\mu}) \Big\} 
\bar u (p_2) \gamma^\nu v(k_2) 
\nonumber\\
&=& 
-(2\pi)^4 \delta^{(4)}(p_1+k_1 - p_2 - k_2)~\aa
\nonumber\\
&\cdot& \Big (
\bar v (k_1) \slb(p_1+k_1) ~u(p_1) \bar u (p_2) (\slp_2+\slk_2) v(k_2) 
\nonumber\\
&+& 
~\bar v (k_1) (\slp_1+\slk_1) u(p_1) \bar u (p_2) \slb(p_1+k_1) ~v(k_2) 
\Big ) 
\nonumber\\
&=& ~ 0 ~.
\nonumber
\ena 
In the same way it is possible to prove the gauge invariance of the second 
graph of Fig.~\ref{psepem}.

\sect{Gauge invariance on bound state.}

The bound state wave function is $\aa$-dependent. This is well-known
for the Schr\"odinger-Coulomb wave function, but is also true for the 
Barbieri-Remiddi solution which reproduces it in the non relativistic limit. 
As a consequence of the non trivial $\aa$-dependence each Feynman graph 
contributes to the energy levels perturbative expansion with a series 
in $\aa$. Also the leading order of this series is not deductable in a 
trivial manner from a vertices counting. As a consequence in a relativistic 
bound state problem it is not possible to verify in a simply way, order 
by order in $\aa$, the gauge invariance.

Although the difficulties to reconstruct sets of gauge invariant 
contributions, since some graphs give series which 
converge faster in $\aa$ in a particular gauge than in an other, 
different gauges are used normally in the energy levels calculation.
Typically binding photons (photons connecting fermion lines 
each other) are calculated in Coulomb gauge while annihilation 
and radiative ones in covariant gauge as the Feynman gauge. 
This not only in different graphs but also, where considered possible, 
in the same Feynman graph.

Such an approach is not free from ambiguities. \\
\begin{figure}[htb]
\vskip 0.8truecm
\makebox[1.2truecm]{\phantom b}
\epsfxsize=11truecm
\epsffile{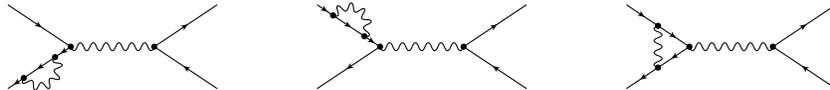}
\vskip 0.3truecm
\caption{{{\small  One-loop vertex corrections to the annihilation 
graph.}}}
\lbl{psann3}
\vskip 0.8truecm
\end{figure}
In Fig.\ref{psann3} there are drawn the vertex corrections at 
the one-photon annihilation graph. After re\-gu\-la\-ri\-za\-tion in order to 
obtain the UV-divergences cancellation it is necessary to consider together 
these three graphs, this means  to use the same gauge for all the photons
(see ~\cite{bur}). If we use a covariant gauge for the radiative photons 
in the first two graphs the third one has to be also calculated in a 
covariant gauge. It follows that there are in any case some binding photons 
(as the no-annihilation photon of the third graph) which will  be calculated 
in a covariant gauge.

In order to avoid these ambiguities and taking in account that the 
apparent simplicity of the Coulomb gauge seems to disappear in 
higher order calculations, a reference calculation of the energy levels 
has to be done in the Feynman gauge (or in an other covariant gauge).
In the following I will show how the individuation of gauge invariant 
set of contributions can be helpful to cancel the low-order spurious 
terms arising in such an {\it ab initio} reference calculation.

The first contribution to the energy levels shift is coming from
the one-photon exchange graph of Fig.\ref{psgamma0}.
\begin{figure}[htb]
\vskip 0.8truecm
\makebox[5.2truecm]{\phantom b}
\epsfxsize=3truecm
\epsffile{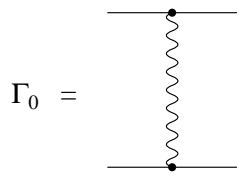}
\vskip 0.3truecm
\caption{{{\small  One-photon exchange graph.}}}
\lbl{psgamma0}
\vskip 0.8truecm
\end{figure}
On the positronium $1S$ state this graph contributes, in the 
Feynman gauge, as ($m=1$):
\eq
\langle \GG_0 \rangle^{Fey}_{1S}  = 
-{1\over 4}\aa^2 -{1 \over 2\pi}\aa^3 \log \aa + {1\over 4 \pi}\aa^3 
-{3\over 16}\aa^4 ~,
\lbl{oneF}
\en
while in Coulomb gauge:
\eq
\langle \GG_0 \rangle^{Cou}_{1S}  = 
-{1\over 4}\aa^2 - {3\over 16}\aa^4 ~.
\lbl{oneC}
\en
As expected, and unlike the $e^+~e^-$ scattering, $\langle \Gamma_0
\rangle$ is not gauge invariant: in Feynman gauge there are some 
contributions (the $\aa^3 \log \aa$ and $\aa^3$ terms) which are not 
in (\ref{oneC}) and in the energy levels itself (it is well-known that 
the first correction to the Balmer's levels is of order $\aa^4$). 
To restore the gauge invariance there must exist some other contributions 
in the energy levels expansion which cancel in Feynman gauge these 
spurious terms. My principal purpose is from now to determine these other 
contributions. I preliminarily give the following general result.

{\it If the zeroth-order kernel $K_c$ is local},
\eq
K_c(\pv,\qv~) = K_c(\pv-\qv~) ~,
\lbl{local}
\en
{\it then}:
\eq
\dd \langle \GG_0  + 2 \sum_{n=1}^\infty I_n \rangle  = 0 ~.
\lbl{giset}
\en
{\it In other words the sum of the contributions to the energy levels coming 
from $\GG_0$ and from the graphs of Fig.\ref{psi} ($\times ~2$) is gauge 
invariant.}
\\
\begin{figure}[htb]
\vskip 0.8truecm
\makebox[1.5truecm]{\phantom b}
\epsfxsize=10.4truecm
\epsffile{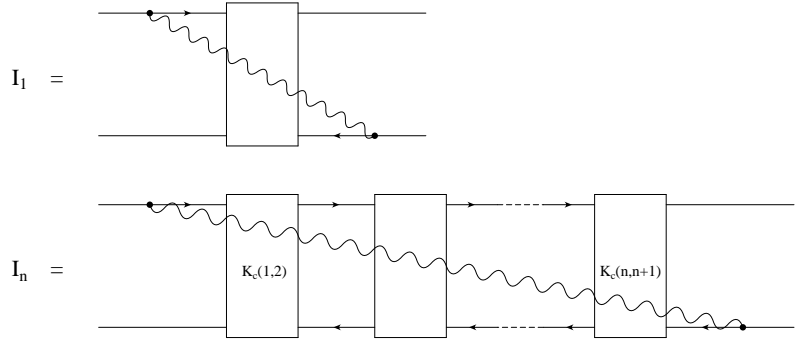}
\vskip 0.3truecm
\caption{{{\small  $n$ kernel $K_c$ crossed by one photon.}}}
\lbl{psi}
\vskip 0.8truecm
\end{figure}

In order to prove (\ref{giset}) first I give the gauge variation of 
$\langle I_n \rangle$:
\eqa
\dd \langle I_n \rangle  &=& \dd \langle K_c G_0 I_n G_0 K_c \rangle  
\nonumber\\
&=& \langle A_n \rangle - \langle A_{n+1} \rangle
+ \langle B_{n+1} \rangle - \langle B_n \rangle ~,
\lbl{i1}
\ena
where the first identity is a consequence of the Bethe-Salpeter equation
for the zeroth-order wave function ($\psi^c = G_0K_c\psi^c$, $G_0$ is the 
two-fermion free propagator),
and the second one is graphically represented in Fig.\ref{psgauge}. 
Fig.\ref{psgauge} also gives the definitions of $A_n$ and $B_n$. \\
\begin{figure}[htb]
\vskip 0.8truecm
\makebox[0.0truecm]{\phantom b}
\epsfxsize=13.4truecm
\epsffile{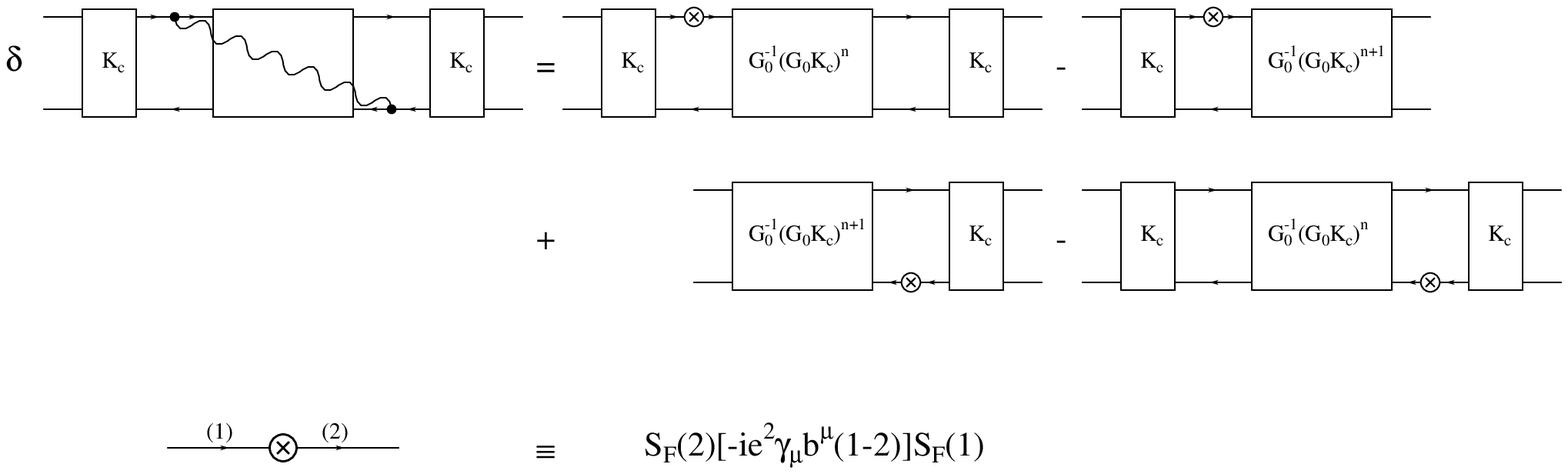}
\vskip 0.3truecm
\caption{{{\small  $ \dd \left(~K_c G_0 I_n G_0 K_c\right) = 
A_nG_0K_c - A_{n+1} + B_{n+1} - K_cG_0B_n $   .}}}
\lbl{psgauge}
\vskip 0.8truecm
\end{figure}
From (\ref{i1}) it follows:
\eq
\dd \sum_{n=1}^\infty \langle I_n \rangle  = 
\langle A_1 \rangle  - \langle B_1 \rangle  ~,
\lbl{i2}
\en
and:
\eq
\dd \langle \GG_0 \rangle = \langle A_0 \rangle
- \langle A_1 \rangle + \langle B_1 \rangle - \langle B_0 \rangle  ~.
\lbl{i3}
\en
It is easy to verify that:
\eqa
\langle A_0 \rangle  &=& \langle B_1 \rangle ~,
\\
\langle B_0 \rangle  &=& \langle A_1 \rangle ~,
\ena
then from (\ref{i2}), (\ref{i3}) one obtains (\ref{giset}).

Similar to the artificial graphs of Fig.\ref{psi} are the graphs
of Fig.\ref{psgamma} which effectively contribute to the 
kernel $K$ of the Bethe-Salpeter equation.  \\
\begin{figure}[htb]
\vskip 0.8truecm
\makebox[2truecm]{\phantom b}
\epsfxsize=9.4truecm
\epsffile{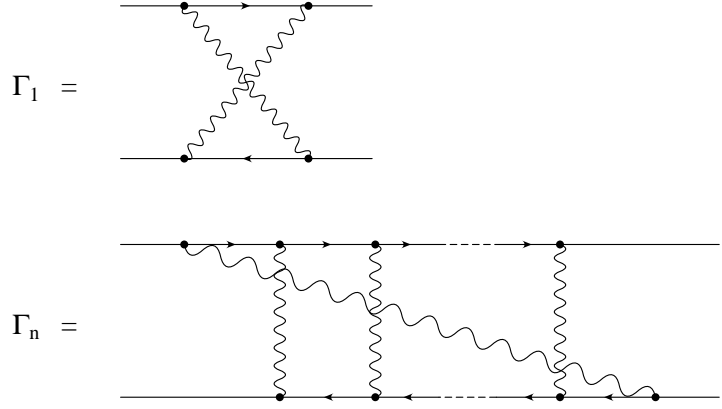}
\vskip 0.3truecm
\caption{{{\small  $n$  ladder photon crossed by one other.}}}
\lbl{psgamma}
\vskip 0.8truecm
\end{figure}
Since,
\eqa
\langle \GG_1 \rangle^{Fey}  &=& 
2 \langle I_1 \rangle^{Fey}   + \langle T \rangle^{Fey}   + O(\aa^4)         ~,
\lbl{g1}\\
\langle \GG_n \rangle^{Fey}    
&=& \langle I_n \rangle^{Fey}   + O(\aa^4) ~~~~~  n >1 ~,
\lbl{gn}
\ena
and up to order $\aa^4$ the Barbieri-Remiddi kernel is local, using 
the result (\ref{giset}), one expects that the spurious terms in (\ref{oneF})
are cancelled by the ones coming from (\ref{g1}) and (\ref{gn}). In fact,
\eqa
2 \langle I_1 \rangle^{Fey}_{1S}  &=& 
{1\over 2\pi}\aa^3\log\aa -{5\over 2\pi}\aa^3
+ {4\over \pi}\log 2 ~\aa^3 + O(\aa^4) ~,
\lbl{two}\\
\sum_{n=2}^\infty \langle I_n \rangle^{Fey}_{1S} &=& 
{9\over 8\pi}\aa^3 -{2\over\pi}\log 2 ~\aa^3 + O(\aa^4) ~.
\lbl{enne}
\ena
The sum of (\ref{two}) and (\ref{enne})  ($\times 2$ taking in account the 
symmetric graphs) cancel completely the $\aa^3$ terms in (\ref{oneF}). 

Up to order $\aa^3$ the term $\langle T \rangle^{Fey}$, which comes 
from $\GG_1$ subtracting from each photon the zeroth-order kernel $K_c$, 
contributes also to $\langle \GG_1 \rangle^{Fey}$ :
\eq
\langle T \rangle^{Fey}_{1S} = {1\over 2\pi}\log 2 ~\aa^3 + O(\aa^4) ~.
\lbl{ttt}
\en
To obtain the cancellation of (\ref{ttt}),  
the graph of Fig.\ref{pskgk}, arising from the second order
perturbations at the energy levels, must be considered; in fact,
\eq
\left \langle ( \GG_0 - K_c) G_0 (\GG_0 -K_c) \right \rangle^{Fey}_{1S}  = 
-{1\over 2\pi}\log 2 ~\aa^3 + O(\aa^4) ~.
\en
\vskip 10truecm
\begin{figure}[htb]
\vskip 0.8truecm
\makebox[4truecm]{\phantom b}
\epsfxsize=5.4truecm
\epsffile{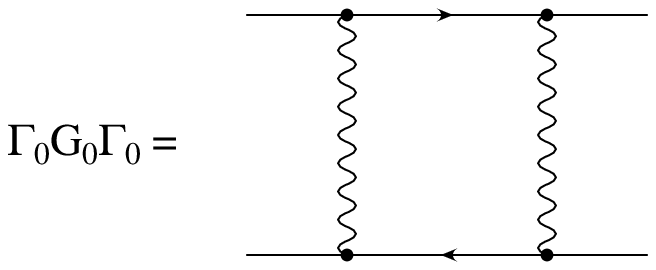}
\vskip 0.3truecm
\caption{{{\small  $\GG_0G_0\GG_0$ graph.}}}
\lbl{pskgk}
\vskip 0.8truecm
\end{figure}

\end{document}